\documentstyle[sprocl]{article}

\input{psfig}

\begin{document}

\title{Extended Quintessence: imprints on the cosmic microwave background 
spectra}

\author{C. Baccigalupi, F. Perrotta}

\address{SISSA/ISAS, Via Beirut 4, 34014 Trieste, Italy, \\
E-mail: bacci@sissa.it, perrotta@sissa.it}

\author{S. Matarrese}

\address{Dipartimento di Fisica `Galileo Galilei', Universit\'a di
Padova,\\ 
and INFN, Sezione di Padova, Via Marzolo 8, 35131 Padova, Italy, \\
E-mail: matarrese@pd.infn.it}

\maketitle\abstracts{We describe the observable features of 
the recently proposed Extended Quintessence scenarios 
on the Cosmic Microwave Background (CMB) anisotropy 
spectra. In this class of models a scalar field $\phi$, assumed to provide 
most of the cosmic energy density today, is non-minimally coupled to the 
Ricci curvature scalar $R$. 
We implement the linear theory of cosmological perturbations 
in scalar tensor gravitational theories to compute CMB 
temperature and polarization spectra. 
All the interesting spectral features are affected: 
on sub-degree angular scales, 
the acoustic peaks change both in amplitude and 
position; on larger scales the low redshift dynamics enhances 
the Integrated Sachs Wolfe effect. 
These results show how the future CMB experiments could give 
information on the vacuum energy as well as on the structure of 
the gravitational Lagrangian term.} 

\section{Introduction}

One of the most interesting novelty in modern cosmology is 
the observational trend for an
accelerating Universe, as suggested by distance measurements 
to type Ia Supernovae \cite{Perlm}. 
These results astonishingly indicate that almost two thirds 
of the energy density today is {\it vacuum} energy. 

It has been thought that this vacuum 
energy could be mimicked by a minimally-coupled scalar field 
\cite{Stain1}, considered as a "Quintessence" (Q). 
The main features of such a vacuum energy component, that could
also allow to distinguish it from a cosmological constant, are 
its time-dependence as well as its capability to develop spatial 
perturbations. 

Theoretically, Quintessence models are attractive, since they 
offer a valid alternative explanation of the smallness of the 
present vacuum energy density instead of the cosmological constant; 
indeed, we must have $| \rho_{vac}| < 10^{-47}$ GeV$^4$ \ today, while
quantum field theories would predict a value for the cosmological
constant which is larger by more than 100 orders 
of magnitude \cite{Lambda}. Instead, the vacuum  
energy associated to the Quintessence is dynamically evolving towards
zero driven by the evolution of the scalar field. 
Furthermore, in the Quintessence scenarios one can select a subclass of
models, which admit "tracking solutions" \cite{Stain1}: here a
given amount of scalar field energy density today can be reached 
starting from a wide set of initial conditions. 

The effects of possible couplings 
of this new cosmological component with the other species 
have been explored in recent works, 
both for what regards matter \cite{carrollame0} and gravity
\cite{ChibaAmeUzan}. 
Here we review some of the results obtained in a recent paper 
\cite{EQ}, for what concerns the effects on the Cosmic Microwave 
Background (CMB) anisotropy: this scenario has been named 
`Extended Quintessence' (EQ), by meaning that 
the scalar field coupled with the Ricci scalar $R$
has been proposed as the Quintessence candidate, in analogy with 
Extended Inflation models \cite{exte}. 

\section{Cosmological dynamics in scalar-tensor theories of gravity}

The action 
$S=\int  d^4 x \sqrt{-g} [ F(\phi)R -
\phi^{; \mu} \phi_{; \mu} -2V( \phi)+ L_{fluid} ]$ 
represents scalar-tensor theories of gravity, 
where $R$ is the Ricci scalar and $L_{fluid}$ 
includes matter and radiation. 

We assume a standard Friedman-Robertson-Walker (FRW) form for the
unperturbed background metric, with signature $(-,+,+,+)$, 
and we restrict ourselves to a spatially 
flat universe. The FRW and Klein Gordon equations are 
\begin{equation} 
\label{FriedmannKleinGordon}
{\cal H}^2=
{a^2 {\rho}_{fluid}\over 3F} +{\dot{\phi}^2\over 6F} + 
+{a^2 V\over 3F} - {{\cal H}\dot{F}\over F}\ ,\ 
\ddot{\phi}+2{\cal H} \dot{\phi}=
{a^2 F_{, \phi}R\over 2} -a^2 V_{,\phi}\ ,
\end{equation}
where the overdot denotes differentiation with respect to the conformal
time $\tau$ and ${\cal H } = {\dot a}/a$.
Furthermore, the continuity equations for the 
individual fluid components are 
$\dot{\rho}_i = -3 {\cal H} ({\rho}_i + p_i)$. 

For what concerns our treatment of the perturbations \cite{EQ}, 
we give here only the very basic concepts. 
A scalar-type metric perturbation in the synchronous gauge is
parameterized as 
\begin{equation}
ds^2=a^2 [-d\tau ^2 + (\delta_{i j }+h_{i j })dx^i dx^j ] \  \ ;
\end{equation}
by linearly perturbing the Einstein and Klein Gordon equations 
above, the equation for the metric perturbing quantities can 
be derived; these equations are linked to the fluid perturbed 
quantities, from any species including $\phi$, 
obeying the perturbed continuity equations. 

Let us define now the gravitational sector of 
the Lagrangian. We require that $F$ has the correct physical 
dimensions of $1/G$. Note that all this fixes the link between 
the value of $F$ today and the Newtonian gravitational constant $G$:
$F_{0}=F(\phi_{0})=1/8\pi G$. Different forms of $F(\phi )$ can be 
considered \cite{EQ}. 
In Induced Gravity (IG) models, that we treat here, 
the gravitational constant is directly 
linked to the scalar field itself, as originally proposed in the 
context of the Brans-Dicke theory: 
\begin{equation}
F(\phi )=\xi\phi^{2}\ ,
\label{IG}
\end{equation}
where $\xi$ is the IG coupling constant. 
Note that solar system experiments already offer 
constraints to the viable values of $\xi$, that may be easily 
obtained by integrating the background equations \cite{EQ}. 
The dynamics of $\phi$ 
is determined by its coupling with $R$, as well as by its 
potential, that is responsible for the vacuum energy today; 
we take the simplest inverse power potential, 
$V(\phi )=M^5/\phi$, where the mass-scale $M$ is fixed by the 
level of energy contribution today from the Quintessence. 
In our integrations, 
we adopt adiabatic initial conditions. 
We require that the present value of $\Omega_{\phi}$ is $0.6$, 
with Cold Dark Matter at 
$\Omega_{CDM}=0.35$, three families of massless neutrinos, 
baryon content $\Omega_{b}=0.05$ and Hubble constant $H_{0}=50$ Km/sec/Mpc; 
the initial kinetic energy of $\phi$ is taken equal to the 
potential one at the initial time $\tau=0$. 

\begin{figure}[t]
\centerline{
\psfig{file=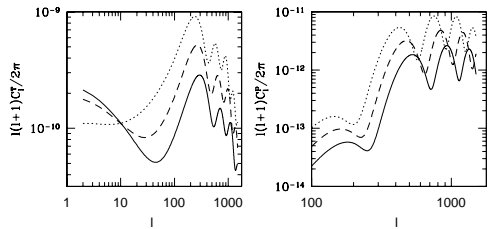,height=3.in,width=3.in}
}
\vskip -4.cm
\caption{Perturbations for IG models for various 
values of $\xi$: 
CMB temperature (left), and polarization (right).
Dotted, dashed, solid lines are for $\xi=0,.01,.02$ 
respectively}
\label{fig}
\end{figure}
%\begin{figure}[t]
%\rule{5cm}{0.2mm}\hfill\rule{5cm}{0.2mm}
%\vskip 2.5cm
%\rule{5cm}{0.2mm}\hfill\rule{5cm}{0.2mm}
%\psfig{figure=filename.ps,height=1.5in}
%\caption{A generalized cactus tree: the confluent
%transfer-matrix $S$ transforms the state function $f(x)$ and   
%$f(z)$ into $f(x)$.  \label{fig:radish}}
%\end{figure}

\section{Effects on the CMB} 

The phenomenology of CMB anisotropies in EQ models 
is rich and possesses distinctive features. 
In Fig.\ref{fig}, 
the effect of increasing $\xi$ on 
the power spectrum of COBE-normalized CMB anisotropies 
is shown. 
The rise of $\xi$ makes substantially three 
effects: the low $\ell$'s region is enhanced, the oscillating 
one attenuated, and the location of the peaks shifted 
to higher multipoles. Let us now explain these effects. 
The first one is due to the integrated Sachs-Wolfe effect, 
arising from the change from matter to Quintessence dominated 
era occurred at low redshifts. This occurs also in ordinary Q models, 
but in EQ this effect is enhanced. 
Indeed, in ordinary Q models the dynamics of $\phi$ is governed by 
its potential; in the present model, one more independent 
dynamical source is the coupling between the Q-field and the Ricci
curvature $R$. The dynamics of 
$\phi$ is boosted by $R$ together with its potential $V$. 
As a consequence, part of the COBE normalization at 
$\ell=10$ is due to the Integrated Sachs-Wolfe effect; 
thus the actual amplitude of the underlying 
scale-invariant perturbation spectrum 
gets reduced. In addition, it can be seen 
\cite{EQ} that the Hubble length was smaller in the past 
in EQ than in Q models. This has the immediate 
consequence that the horizon crossing of a given 
cosmological scale is delayed, making 
the amplitude the acoustic oscillations slightly 
decreasing since the matter content at decoupling is 
increased. 

Finally, note how the location of the acoustic peaks 
in term of the multipole $\ell$ at which the oscillation occurs, 
is shifted to the right. Again, the reason is the time dependence 
of the Hubble length, which at decoupling 
subtended a smaller angle on the sky. It can indeed 
verified that the ratio of the peak multipoles in 
Fig.\ref{fig} coincides numerically with the 
the ratio of the values of the Hubble lengths at decoupling 
in EQ and Q models \cite{EQ}. 

We have used here values of $\xi$ large in order to clearly 
show the CMB effects. 
It can be seen these values do not satisfy the solar system 
experimental constraints; however, a smaller $\xi$ produces 
the same spectral features, reduced but still 
detectable by the future 
generation of CMB experiments, able 
to bring the accuracy on the CMB power spectrum 
at percent level up to $\ell\simeq 1000$ \cite{CMBFUTURE}.

\end{document}